# Broken Symmetry Approach and Chemical Susceptibility of Carbon Nanotubes

**ELENA F. SHEKA**[1], **LEONID A. CHERNOZATONSKII**[2]

[1]*Peoples' Friendship University of the Russian Federation, 117198 Moscow, Russia.*
 sheka@icp.ac.ru

[2]*Institute of Biochemical Physics, Russian Academy of Sciences, 119991 Moscow, Russia.*
 cherno@sky.chph.ras.ru

**ABSTRACT**: Constituting a part of odd electrons that are excluded from the covalent bonding, *effectively unpaired electrons* are posed by the singlet instability of the single-determinant broken spin-symmetry unrestricted Hartree-Fock (UBS HF) SCF solution. A correct determination of the total number of effectively unpaired electrons $N_D$ and its fraction on each atom $N_{DA}$ is well provided by the UBS HF solution. The $N_{DA}$ value is offered to be a quantifier of atomic chemical susceptibility (or equivalently, reactivity) thus highlighting targets that are the most favorable for addition reactions of any type. The approach is illustrated for two families involving fragments of arm-chair (n,n) and zigzag (m,0) single-walled nanotubes differing by length and end structure. Short and long tubes as well as tubes with capped end and open end, in the latter case, both hydrogen terminated and empty, are considered. Algorithms of the quantitative description of any length tubes are suggested.

## 1. Introduction

A growing number of physical, chemical, and biomedical applications of carbon nanotubes (CNTs) require chemical modification of the latter to make them more amenable to rational and predictable manipulation [1]. Two main strategies concerning non-covalent [2] and covalent [3] functionalizations have been elaborated and have gained large practical success and achievements (see comprehensive reviews [4-11]). However, controllable functionalizing techniques which can provide CNTs tailoring in a determinable manner are far from completing that still requires more investigation to elucidate the nature and locality of covalently attached moieties. Due to extreme complication of native CNTs as well as difficulties in their separation and individual studying, a considerable load in looking for answers is put on shoulders of computational approaches. Starting in [5], a number of calculations have been performed [12-21].

Not far long ago the main concept of the connection between the electronic structure and chemical reactivity of CNTs was based on the Haddon approach [22] that attributed the reactivity of fullerenes to the strain engendered by their spherical geometry as reflected in pyramidalization angles of carbon atoms. Just as in the case of fullerenes, curvature-induced pyramidalization and misalighnment of the $\pi$-orbitals of the carbon atoms [23-25] induces a local strain in a defect-free CNT. This concept made allowance for explaining the difference between the reactivity of a single-wall CNT (SWCNT) endcap and sidewall in favor of the former, and gave a simple explanation of the reactivity increase while the SWCNT diameter decreases. Further development of the approach turned out to be useful for concerning the reactivity of the convex and concave sidewalls of SWCNTs towards addition reactions [12, 13].

The Haddon approach was quite productive but basically empirical while more sophisticated theoretical approaches were needed. Various DFT techniques actively explored during the last time with respect to the chemical functionalization of CNTs, to name a few [16-21], were a natural response to the requirement. However, a lot of severe limitations are usually implied when using the technique. This concerns 1) the aromaticity concept attributed to the electronic structure of the tubes that results in the closed-shell approximation for wave functions; 2) the periodic boundary conditions along the tubes that restricts the tube area consideration to sidewall only leaving the tube ends, the most active spots on the tube, as it will be shown below, outside the consideration; 3) the absence of atomically matched characteristics that could exhibit the chemical reactivity of the tube atoms; 4) a *post factum* character of simulations that are aimed at obtaining the energy and structural data consistent with experimental findings by adapting the functional used. The most crucial is the first assumption since CNTs belong to a particular class of species with weakly interacting odd electrons. The term was introduced in organic chemistry when describing the electronic structure of diradicals [26, 27] and naturally covers terming "magnetic electrons" [28] as well as "open-shell molecule electrons' [29]. With respect to CNTs, it manifests that the number of valence electrons of each carbon atom is larger by one than that of interatomic bonds it forms. When the problem has been considered in the framework of the aromaticity concept, this means a strong coupling between the electrons so that odd electrons are covalently coupled in pairs similarly to π electrons of benzene molecule.

The approach generalization for systems with weak coupling that causes the appearance of nearly degenerate spin states, ultimately requires taking into account the electrons correlation and passing to computational schemes that involve full configurational interaction (CI). However, the traditional complete-active-space-self-consistent-field (CASSCF) methods that deal correctly with two-odd electrons systems such as diradicals and some dinuclear magnetic complexes (see a comprehend review [30]), cannot handle systems with larger number of odd electrons due to a huge number of configurations generated in the active space of the system. To imagine arising severe technical difficulties it is worthwhile to cite the paper [30], "If there are '$m$' singly occupied orbitals on each of '$n$' identical centers, then $2^{mn}$ Slater determinants can be formed by assigning spins up or down to each of the '$nm$' orbitals". So that no CASSCF type approach, including a simplified embedded-cluster CASSCF supplemented by the complete active space second-order perturbation theory (CASPT2) [31], seems feasible for many-odd electron systems such as fullerenes, CNTs, and graphene, and addressing single-determinant approaches seems to be the only alternative.

Open-shell unrestricted broken spin-symmetry (UBS) approach suggested by Noodleman [32] is the most known among the latter. The approach is well elaborated for both wave-function and electron-density QCh methodologies, based on unrestricted single-determinant Hartree-Fock scheme [33] and spin-dependent local-spin-density-approximated (LSDA) DFT that, in its turn, is based on the Kohn-Sham single Slater determinant procedure [34]. In what follows we shall refer to them as UBS HF and UBS DFT, respectively.

The UBS approach main problem concerns spin-contamination of calculation results. The contamination contribution depends on the interaction between odd electrons and increases considerably when the interaction decreases. However, if this evident fact is willingly accepted for UBS HF and often is addressed as the approach, convincing disadvantage, a widely spread illusion exists that spin-polarized (UBS) DFT approach provides obtaining pure spin states. In reality, as shown in [29], it is not so and UBS DFT solution is spin-contaminated as well due to single-determinant character of wave functions that form electron density. Ignoring this fact leads to wrong conclusions such as the prediction of unreal magnetic properties of graphene ribbons (see a detailed discussion of the problem in [35].

The main problem concerning the interpretation of UBS results in view of their relevance to physical and chemical reality consists in mapping between the eigenvalues and eigenfunctions of exact and model spin Hamiltonians. While the implementation of UBS HF approach, both *ab initio* and semiempirical, is quite unique and the wished mapping is quite straightforward, this cannot be said about calculation schemes based on LSDA DFT [29]. This is caused by a fundamental problem of DFT concerning the total spin. For long ago has been known that DFT cannot be directly applied to calculation of the spin and space multiplet structure (a comprehensive analysis of problems existing in the present formulation of DFT is performed in [36]). A number of special procedures, that all are beyond the pure DFT scope [37], are suggested to overcome the difficulty. The procedures differ by computation schemes as well as by obtained results so that UBS DFT is method-dependent [36-38]. This greatly complicates a comparison of data obtained by different computational versions and places severe obstacles in computational experiments aimed at obtaining quantitative-structure-activity-relationships (QSARs) attaching them to a particular computational scheme, therewith without a complete sure in the results reliability. This is one reason why the preference of UBS HF is obvious if QSARs are the main goal of the study. The other reason is connected with a complete theoretical transparency of UBS HF scheme when any quantity, including total spin, has a clear physical meaning and a proper description. However, it should be kept in mind that UBS HF is only an approximation to the exact solution of the problem so that a thorough analysis should be done attributing obtained results to physical or chemical reality. Fortunately, there is a well elaborated way providing such mapping that will be discussed in the next Section.

The current study presents a platform for the UBS HF application to quantifying QSARs related to chemical reactivity of single-walled CNTs (SWCNTs). A large set of fragments related to armchair (n,n) (n=4, 5, 6, 8, 9, 10) and zigzag (m,0) (m=8, 10, 12) SWCNTs differing by length and end structure, namely, short and long as well as those with endcaps and open ends, in the latter case, both hydrogen terminated and empty is studied. The results occurred to be quantitatively similar both for the two families as a whole and for the family members within each family. The main result concerns disclosing a quantified chemical susceptibility (or equivalently, reactivity) of the tubes atoms that allowed dividing the tubes space into a small number of characteristic regions related to differently configured ends, sidewall and defects. The finding allows for suggesting a unique algorithm for the quantitative description of the atomically matched chemical susceptibility of SWCNTs of any length and any end composition and gives a clear vision for experimentalists and engineers how to proceed with chemical covalent functionalization of the SWCNTs aiming at obtaining wished chemical compositions.

## 2. Broken Symmetry Approach

UBS HF and UBS DFT find favor over close-shell ones because the "correlation effects" by Löwdin [39] are automatically introduced at least through the different-orbital-different-spin understanding. Both starting approaches, UHF and LSDA correspond to a state with a definite value of the spin projection $S_z$, but does not, in general, correspond to the state with a definite value of the total spin S, so that

$$\hat{S}_z \psi^{UHF} = S_z \psi^{UHF} \text{ and } \hat{S}_z \psi^{LSDA} = S_z \psi^{LSDA}$$
$$\text{but } \hat{S}^2 \psi^{UHF} \neq S(S+1)\psi^{UHF} \text{ and } \hat{S}^2 \psi^{LSDA} \neq S(S+1)\psi^{LSDA}. \qquad (1)$$

As a result, both solutions are spin-contaminated if only all electron spins either are not coupled that leads to S=0 or have not one direction so that S=$S_{max}$. The contamination

$$C = \langle \hat{S}^2 \rangle - S(S+1) \qquad (2)$$

is often substantial.

The inequalities in (1) result in a symmetry dilemma by Löwdin [40] that is expressed as asymmetric electron densities of UHF solution and asymmetric LSDA Hamiltonian with different exchange-correlation potentials for spin-up and spin-down orbitals. The asymmetry produces the appearance of spin density in the general form [41] (see the latest discussion of the problem in [30, 42]

$$D(r|r') = 2\rho(r|r') - \int \rho(r|r'')\rho(r''|r')dr''. \tag{3}$$

In the case of UHF approach $D(r|r')$ is determined as

$$DS = 2PS - (PS)^2 \tag{4}$$

Here $P = P^\alpha + P^\beta$ is the density matrix and $S$ is the orbital overlap matrix (α and β mark different spin directions). If the UBS HF computations are realized via *NDDO* approximation, a zero overlap of orbitals results in $S = I$, where $I$ is the identity matrix. The spin density matrix $D$ then assumes the form [43]

$$D = \left(P^\alpha - P^\beta\right)^2. \tag{5}$$

The elements of the density matrices $P_{ij}^{\alpha(\beta)}$ can be written in terms of eigenvectors of the UBS HF solution $C_{ik}$

$$P_{ij}^{\alpha(\beta)} = \sum_k^{N^{\alpha(\beta)}} C_{ik}^{\alpha(\beta)} * C_{jk}^{\alpha(\beta)}, \tag{6}$$

where $N^\alpha$ and $N^\beta$ are the numbers of electrons with spins α and β, respectively. These explicit equations are the consequence of Ψ-based character of UHF. Since the corresponding coordinate wave functions are subordinated to the definite permutation symmetry, each value of spin $S$ corresponds to a definite expectation value of energy [37].

Oppositely, the electron density ρ is invariant to the permutation symmetry [37]. That is why spin density $D(r|r')$ of LSDA depends on spin-dependent exchange and correlation potentials and cannot be expressed analytically. Since the exchange-correlation composition deviates from one method to the other, the spin density is not fixed and deviates alongside with the composition.

Leaving aside the problem of the pure-spin state extraction from spin-contaminated unrestricted solutions that is complicated differently for UHF and UBS DFT, let us discuss a pragmatic usefulness provided with UBS solutions. Obviously, spin-contamination $C$ and spin density $D(r|r')$ exhibit incompleteness of taking correlation effects into account. The latter, governed by the Fock exchange only, is evidently the largest for UHF. As for UBS DFT, there are additional possibilities concerning functionals in use: from the Fock exchange only functional that causes a severe underestimation of the correlation effects to local density approximation (LDA) that provides a huge overestimation, so mixing the Fock and LDA exchange can provide any beforehand decided (method-dependent) result [29, 38]. Oppositely to the fact, UBS HF solution is strictly standard. At the same time, the solution retains a transparent single-determinant description, restricted version of which led the foundation of the modern chemistry language, due to which a question arises if there is a possibility "to translate" the correlation incompleteness of UBS HF in convenient terms so that it will become practically useful if properly designed.

A suggestion to use the correlation incompleteness of the UBS solution for practical goals is usually connected with the Noodleman publication [32]. Since then started the term "broken symmetry approach". However, the Noodleman consideration was restricted to obtaining exchange integral $J$ that is directly related to the energy difference of Heisenberg Hamiltonian eigenstates and is of big importance for considering magnetic properties. The parameter provides as well a correct calculation of the energy of pure spin states. However, the first addressing to UBS solution in view of its use for practical chemistry was made in [41]. The authors showed that the extra spin density of UBS solution $D(r|r')$ can be interpreted in terms of *effectively unpaired electrons* (EUPEs) that constitute a part of valence electrons withdrawn from the covalent bonding. Within the framework of UBS HF the quantity can be explicitly quantified in the form [43]

$$N_{DA} = \sum_{i \in A} \sum_{B=1}^{NAT} \sum_{j \in B} D_{ij} \qquad (7)$$

and

$$N_D = \sum_A N_{DA} = \sum_{i,j=1}^{NORBS} D_{ij}, \qquad (8)$$

where $N_{DA}$ determines the EUPEs number on atom $A$ while $N_D$ constitutes the total EUPEs number. The summation is performed over the number of atoms in the first case and over all atomic orbitals in the second.

Firstly applied to fullerenes [43, 44], $N_{DA}$ was shown to disclose chemical activity of individual molecule atoms that was termed as *atomic chemical susceptibility* (ACS). Similarly referred to, $N_D$ can be called as *molecular chemical susceptibility* (MCS). Atomically matched $N_{DA}$ is a quantifier thus highlighting targets that are the most favorable for addition reactions of any type. Practical realization of computational synthesis of the fullerene derivatives proved a great importance of the two quantities in describing chemical reactions and their products.

Oppositely to UBS HF, UBS DFT does not suggest any quantity similar to either $N_D$ or $N_{DA}$. This might be a reason why UBS DFT calculations are rather scarce. Concerning CNTs, the only UBS DFT calculation has been known so far [45] that thoroughly analyzed the singlet state of cyclacenes and short zigzag SWCNTs. It was shown that the energy of UBS DFT singlet state was lower than that of restricted DFT singlet that points to an open-shell character of the singlet state. The finding is of a great importance exhibiting invalidness of aromaticity-based concept of DFT for a quantitative description of the electronic states of CNTs.

## 3. Results

Firstly applied to two fragments of (4,4) defect-free and (4,4) (5-7) defect SWCNT as example [46], UBS HF approach has demonstrated that $N_{DA}$ values appear and follow in synchronism with the excess of C-C distance that separates two odd electrons over a limit value $R_{cov} \cong 1.395\text{Å}$. The limit corresponds to the largest C-C bond length at which a complete covalent bonding of the electrons occurs. Therefore, the $N_{DA}$ map which describes the tube atoms ACS is tightly connected with the tube structure thus highlighting the distribution of the C-C bond length excess over the tube. To prove the statement, an extended set of fragments of (n,n) and (m,0) SWCNTs have been considered in the current study. Calculations were performed in semiempirical approximation by using AM1 version of the CLUSTER-Z1 program [47] that provides stable UHF calculations in singlet states supplemented by the calculations of all above-mentioned quantitative characteristics related to EUPEs. A particular attention was given to initial guess to both $P^\alpha$ and $P^\beta$ matrices when performing calculations in the singlet state. An obvious choice in setting these matrices to zero and using $H^{core}$ as an initial guess to both matrices is followed by producing identical orbitals for α and β spins [48]. To avoid the complication, the relevant matrices were obtained in due course of one-point-geometry calculations in the triplet state (or any other high spin state) and then were input as the initial guess for the singlet calculations.

### 3.1. (4,4) SWCNTs

Equilibrium structures of the studied fragments were obtained in due course of the structure optimization when seeking the energy minimum and are presented in Fig.1 and Fig.2. Numbered from NT1 to NT9 the fragments form three groups. The first group (Fig.1) involves three fragments related to short defect-free (4,4) SWCNTs differing by the end structure. One end of all fragments is capped while the other is either open but hydrogen terminated (NT1) and empty (NT2), or capped (NT3). Attached hydrogens are not only service terminators but present a real hydrogenation of the tube open end. The collection is added by a defect (4,4) SWCNT (NT4) with a Stone-Wales pentagon-heptagon pair. The second group (Fig.2) joins fragments NT5 and NT6 that present capped-end & open-end (4,4) SWCNTs but differing by the termination of the open end. They differ from fragments NT1 and NT3 by four atom rows elongation. The third group (Fig.2) combines NT7, NT8, and NT9 fragments of (4,4) SWCNTs with both open ends but differing by hydrogen termination. The fragment sidewall is longer by one atom row comparatively with that of the second group. Taking together, the fragments allow for considering the following structure effects on the $N_{DA}$ distribution caused by i) endcaping; ii) terminating or emptying open ends; iii) introducing a pair of pentagon-heptagon defects; iv) the tube sidewall elongation.

**Fragments of group 1.**

Eight hydrogen atoms terminate the NT1 open end. The atoms are numbered from the cap towards the open end following their symmetrical disposition with respect to both $C_2$ axis and the symmetry plane passing through the axis normally to the paper sheet. The numbering of atom rows proceeds in the opposite direction. This enumeration will be retained in what follows when presenting calculation data for other fragments.

As shown previously [46], the dominating majority of C-C bonds of the tube are longer than the limit $R_{cov}$ =1.395Å, so that the large total number of EUPEs for the tube $N_D$= 32.38 does not look strange. The distribution of these electrons over the atoms forms the ACS $N_{DA}$ map that is shown in Fig.3(a). As seen from the figure, the map shape supports $C_{2v}$ symmetry of the atom arrangement. According to the map, the tube can be divided into three regions. The first is related to the cap with adjacent atoms and covers rows 9-14. The second concerns mainly the tube sidewall and covers rows 4-8. The third refers to the open end terminated by hydrogen atoms and covers rows 1-3. The biggest non-uniformity of the $N_{DA}$ distribution is characteristic for the cap region. One should draw attention that the largest $N_{DA}$ values belong to atoms that are characterized by the largest C-C bond length and vice versa. As for the sidewall region, the $N_{DA}$ distribution is practically uniform with the $N_{DA}$ value scatter not bigger than 0.5% that is consistent with quite uniform distribution of C-C bond lengths as well. The $N_{DA}$ distribution in the end region is significantly affected by the hydrogenation. But in this region there is a direct correlation between the $N_{DA}$ values and lengths of C-C bonds as well. Therefore, the C-C bond length is actually a controlling factor in the distribution of the EUPEs density over atoms.

The curve with dots in Fig. 3(a) presents the free valency of the tube atoms calculated in accordance with the relation

$$V_A^{free} = N_{val}^A - \sum_{B \neq A} K_{AB} \tag{9}$$

where $N_{val}^A$ is the number of valence electrons of atom $A$ and $\sum_{B \neq A} K_{AB}$ is the generalized bond index

$$K_{AB} = |P_{AB}|^2 + |Sp_{AB}|^2. \tag{10}$$

Here, the first term is the Wiberg bond index [49] and the second is determined by the spin density matrix (see [50] for details). The excellent agreement of the two values shows that the $N_{DA}$ map does exhibit a quantitative measure of free valency or ACS. From this viewpoint, the tube cap is the most reactive part, while the tube sidewall is more passive with ill-pronounced selectivity along the tube.

Removing hydrogen atoms at the tube open end, we obtain the $N_{DA}$ map of NT2 fragment shown in Fig. 3(b). A tremendous contribution of end atoms obviously dominates on the map. It is due to the fact that the ethylene-like (*2n*) C-C bonds [44] are replaced by the acetylene-like (*3n*) bonds at the tube end where each atom has not 1 but 2 odd electrons. The transformation naturally results in increasing the total EUPEs number $N_D$ from 32.38 up to 39.59. The injection of additional EUPEs disturbs the $N_{DA}$ map of the hydrogen-terminated tube shown in the figure by bars quite considerably. Important to note that changes occur not only in the vicinity of the open end within rows *2* and *3* that is quite reasonable, but influences the opposite cap end (rows *14-11*). Practically no changes occur along the tube sidewall which seems to serve as a peculiar resonator for the electron conjugation. Addressing the chemical activity of the tube, it becomes evident a dominant activity of the empty end atoms.

When the other end is capped, the fragment NT3 ($N_D$=33.35) becomes highly symmetrical ($D_{2h}$) that is reflected in its $N_{DA}$ map shown in Fig. 3(c). The $N_{DA}$ distribution of the fragment is specularly symmetrical with respect to atoms of the middle row *8*. At the same time, as seen from the figure, the addition of the second cap results in no changing in the distribution which is characteristic for the one-cap fragment NT1.

The defect introduction in the NT4 fragment was purposely made above the symmetry axis to distinctly violate $C_{2v}$ symmetry of the perfect tube. The tube symmetry becomes $C_s$. As seen in Fig.1, the defects are spread over rows 5-7 centering at row 6. A comparative analysis of the $N_{DA}$ maps of perfect and deformed tubes presented in Fig. 4 shows that, actually, the main alterations concern atoms of these rows. The remainder part of the sidewall region is less affected. As for the cap and open end regions of the tube, the alterations also are rather weak. And nevertheless, whilst small, the changes caused by the defects touch upon all atoms of the *12*-row tube.

A significance of the open end termination can be traced through charge characteristics of the fragments as well. In the series NT1, NT2, NT3 dipole moments of fragments are 9.93, 0.44, and 0.03 D, respectively. A drastic drop of the value when terminating hydrogen atoms are removed is evidently connected with charge distribution over tube atoms. Fig. 5 presents that for the three fragments. Bars in the figure show the charge distribution along NT1 fragment with atom enumeration that coincides with that of the $N_{DA}$ map in Fig. 3(a). A $C_{2v}$ symmetry pattern of the distribution is clearly seen. As seen from the figure, attached hydrogens as well as adjacent carbon atoms provide the main contribution to the distribution. The influence of terminators is still sensed at another two atomic layers towards the tube cap. The remainder tube sidewall is practically uncharged, and a weak charging of atoms appears again only at the endcap. The acquired total negative charge is fully compensated by positively charged hydrogen atoms. However, significantly charged "tail" of the fragment provides a high value of the dipole moment.

When hydrogen atoms are removed the charge map is reconstructed. First, it concerns end carbon atoms whose charge becomes much less but the charging area is still large covering rows *1-3*. Besides, the charge redistribution is seen in the cap region as well while along the tube sidewall changes are small. And again, as in the case of the $N_{DA}$ distribution, one can speak about a resonant character of the disturbance transfer along the tube sidewall. One might suggest that this very charge transfer along the tube results in so significant decreasing of the dipole moment value in spite of a still considerable charging of the tube open end.

Replacing the open end by the second cap does not violate the charge distribution in the region of the first cap and simply duplicates it in the region of the second cap just making the total tube $N_{DA}$ map specularly symmetrical with respect to the middle row of atoms. Highly symmetric and low by value, the charge distribution produces exclusively small dipole moment.

**Fragments of group 2**

$N_{DA}$ maps given for fragments NT5 and NT6 in Fig. 6 show the following. Like for members of group 1 considered in the previous Subsection, the $N_{DA}$ map of the fragments consists of three regions related to the cap end on the left, open end on the right, and an extended tube sidewall. As for both end regions, the data for elongated tube fully reproduce those for a shorter tube not depending on if the open end is either hydrogen terminated (Fig. 6(a)) or empty (Fig. 6(b)). A similar behavior of the map should be expected if the open end is capped. The only difference in the distributions related to shorter and longer fragments consists in expending a homogeneous part of the distributions related to the tube sidewalls. Therefore, the $N_{DA}$ map may be presented as consisting of three fragments MapI, MapII, and MapIII. In the case of two cap ends, MapIII should be replaced by MapI. As seen from Fig. 6, MapI covers region of first (and/or last) 37-40 atoms. MapIII is spread over 24 last atoms. MapII takes the remainder space which depends on the tube length.

**Fragments of group 3**

Group 3 covers fragments with both open ends and is consistent by length with fragments of group 2. Similarity in the tube length allows for revealing changing caused by replacing one or two cap ends by open ends.

Fig. 7(a) presents a comparative view on $N_{DA}$ maps of fragments NT5 and NT7 that have similar right open ends terminated by hydrogen and different left ends presented by the cap in the case of NT5 and by hydrogen terminated open end in NT7. A comparison shows that if three-region pattern described above is characteristic for capped NT5 fragment, this can not be said about NT7. The $N_{DA}$ map in the case is specularly symmetrical with respect to the middle row but is quite inhomogeneous with a peculiar step-like character reaching minimum at the tube center. It should be noted nevertheless that the step character related to the last three rows on the right end is the same as for the NT5 open end.

Removing hydrogen atoms on the right end of the tube brings back the $N_{DA}$ map of fragment NT8 to a three-region form described in the previous Subsection (see Fig 7(b)) but expressed as MapIII+MapII+MapIII$^*$. MapII and MapIII have the same shape as previously while MapIII$^*$ concerns 24 last atoms after removing hydrogen terminators. When the remainder eight terminators are removed on the left end, the $N_{DA}$ map of fragment NT9 acquires a new three-region form expressed as MapIII$^{**}$+MapII+MapIII$^*$. As a whole the map becomes almost specularly symmetrical with respect to the middle row of atoms while MapIII$^{**}$ somewhat differs from MapIII$^*$. When fragment NT9 is formed from NT7 avoiding the NT8 stage, MapIII$^{**}$ is fully identical to MapIII$^*$.

**3.2. (n,n) and (m,0) SWCNTs**

**(n,n) SWCNTs**

The obtained results show that the three-region pattern of the $N_{DA}$ maps is generally supported for all studied fragments, whilst not in the same manner. Thus, the graphical views of the $N_{DA}$ maps for (5,5); (6,6); (7,7); and (8,8) members are quite similar to those of the (4,4) set whilst naturally differing numerically. The corresponding $N_{DA}$ values are summarized in Table 1. When the deviation is big, the related interval of the value changes is shown. Analysis of the data given in the table makes allowance for exhibiting the dependence of the ACS on the SWCNT diameter. As seen from the table, the dependence is different in different regions. Thus, ACS of the H-terminated ends does not show any dependence at all. Carbon atoms of empty ends form two groups, equal by number, characterized by two $N_{DA}$ values. The bigger value of 1.16 does not change when the diameters increases while the lower one drops from 1.06 to 0.67 when passing from (4,4) SWCNT to (5,5) one and then remains unchanged. The side wall ACS decreases gradually when the diameter increases and approaches to a steady value when the diameter grows further. Qualitatively, the ACS behavior in this region looks like that expected on the basis of the Haddon approach. Obviously, the $N_{DA}$ values will approach the ones of a graphene sheet when the diameter further increases. As for the cap region, a large scattering of the ACS values as well as their peculiar change reflect a severe reconstruction of the C-C bond net in the region adapting it to a minimal stress of the tube body as a whole and its cap end, in particular. From this viewpoint, small $N_{DA}$ values at the cap of the (5,5) tube is obvious. The tube symmetry is $C_{5v}$ and the cap structure is comfortably centered around pentagon. The C-C bonds are the least stressed and lengthened that causes the least $N_{DA}$ values.

As seen from Table 1, oppositely to the previous case, the scattering of the $N_{DA}$ values of tubes (9,9) and (10,10) is significant. Particularly, small $N_{DA}$ values in the sidewall and open H-terminated regions should be mentioned. We attribute this feature to a great stress of the tube bodies that is obviously seen in Fig.8. As follows, one cannot expect in this case regular C-C bond lengths distribution not only in the cap region, but on sidewall and end atoms as well. The observed decrease of the $N_{DA}$ values is connected with the C-C bond shortening that reflects the body stress. At the same time, the largest values in the sidewall and H-terminated end region are similar to those of the previous group and clearly show their saturation under the tube diameter growing. Similar characteristics might be expected for the tube of larger diameter as well.

In spite of a large scattering of the $N_{DA}$ data in the sidewall region, their distribution retains a regular view. The matter is that the scattering takes place within an individual row of the sidewall while repeating regularly over rows. The $N_{DA}$ distribution within one row of the studied tubes is shown in Fig.9. As seen from the figure, it looks quite similar for both tubes as well for other rows of the two ones.

**(m,0) SWCNTs**

While (n,n) SWCNTs all are semiconductive, the (m,0) SWCNTs family consists of both semiconductive and metallic (really, semimetallic) tubes depending on whether *m* is divisible by 3. Therefore, a series of (8,0), (10,0) (two of different length), and (12,0) fragments contains two semiconductive and one metallic tube and offers two new aspects for a comparative study. The latter concerns changes of $N_{DA}$ maps caused by (1) replacing armchair open ends of (n,n) tubes by zigzag ones for (m,0) family and (2) changing the tube conductivity within (m,0) family.

In response to the first inquiry, the performed calculations have revealed a large similarity in the behavior of (n,n) and (m,0) fragments. A comparative view on the $N_{DA}$ maps of (4,4) and (8,0) fragments that both are semiconductive and close by diameter is presented in Figs.10a and 10b. As seen from the figure, $N_{DA}$ maps of the two tubes are both qualitatively and quantitatively similar. First, the maps consist of three parts related to variable $N_{DA}$ values at capped and open ends and of rather homogeneous region of the $N_{DA}$ values along the tube sidewalls. In both cases open empty ends are the most chemically reactive. Following these places in activity are endcaps and sidewall. Second, not only the shape of $N_{DA}$ maps but their numerical plottings are practically the same for endcaps and sidewalls (Table 1). The only difference concerns zigzag open ends, both H-terminated and empty, that exhibit higher reactivity in comparison with those of armchair ends. These regularities retain to a great extent for SWCNTs of bigger diameter: compare the data in Table 1 for (6,6) and (10,0) as well as for (7,7) and (12,0) tubes.

The latter is of particular interest since (7,7) SWCNT is semiconductive while (12,0) is metallic. More detailed a comparison between maps of the tubes of different conductivity is visualized in Fig. 10c for (10,0) and (12,0) SWCNTs belonging to the same family. As seen from the figure, plottings for both tubes are well similar by both shape and numbers and do not exhibit any remarkable characteristic difference that might be related to changing in the tubes conductivity. Therefore, the chemical reactivity of atoms of both semiconductive and metallic SWCNTs is comparable that does not cause different behavior of the tubes with respect to similar chemical reactions. This is consistent with the reality when one faces the problem of using sidewall covalent chemistry for tube separation [51]. Partial lucks in achieving the goals of distinguishing semiconductive and metallic tubes by using chemical modification seem to be connected with peculiarities of intermolecular interaction between the tube and corresponding addends caused by donor-acceptor interaction.

## 4. Discussion and Conclusive Remarks

### 4.1. GENERAL VIEW ON SWCNTs CHEMICAL REACTIVITY

Exhibited peculiarities of the obtained SWCNT ACS ($N_{DA}$) maps allow for making the following conclusions concerning additional reactions to be expected.

1. The space of chemical reactivity of any SWCNT coincides with its coordinate space whilst different for particular structure elements. This simultaneously both complicates and facilitates chemical reactions involving the tubes depending on a particular reaction goal. A summarized view on chemical reactivity of the tubes, presented in details by $N_{DA}$ maps in Figs. 3, 4, 6, 7, and 10 is given in Table 1. Shifting the activity to either cap or empty end regions depending on the tube shape and composition may be used in looking for practical ways of the tube chemical covalent modification.

2. Local additions of short-length addends (involving individual atoms, simple radical and so forth) to any SWCNT are the most favorable at open empty ends, both armchair and zigzag ones, the latter more effective. Following these places in activity are endcaps, defects in the tube sidewall, and sidewall itself.

3. Chemical reactivity of SWCNTs open but chemically terminated ends exceeds that of sidewall only for SWCNTs with zigzag ends.

4. Single local addition of long-length addends (polymers) will follow the same

rules while wrapping along the SWCNT sidewall will be the most favorable due to a large number of local contacts on the way.

5. Chemical contacts of SWCNTs with spatially extended reagents (surfaces of crystalline and amorphous solids, graphene sheets, etc) can occur in three ways when the tube is oriented either normally or parallel to the surface and when the latter (graphene) acts as a cutting blade [52].

6. Addition reactions with participation of MWCHTs will proceed differently depending on the target atoms involved. If empty open ends of the tubes are main targets, the reaction will proceed as if one deals with an ensemble of individual SWCNTs. If sidewall becomes the main target the reaction output will depend on accessibility of inner tubes sidewall additionally to the outer one.

## 4.2. COMPARISON WITH EXPERIMENT

Lack of experimental data for individual SWCNTs complicates the exact verification of the obtained data. However, there are some generalized observations that can be discussed in view of the performed computations.

1. Studies performed by Yates et al. [53] and Yates and Smalley et al. [54] can be attributed to investigation of oxidatively cut SWCNT caps of ~1 nm in diameter thus revealing empty open ends interacting with primarily esters and quinones. Such oxidized SWCNTs have been assembled on a number of surfaces, including silver [55], highly oriented pyrolytic graphite [56], and silicon [57]. High coverage densities and orientation normal to the surface have been shown; the latter being suggestive of higher degrees of functionalization at the nanotubes ends. The findings well correlate with calculated exclusive reactivity of the SWCNT empty-open-ends.

2. A detailed study of diameter-dependent oxidative stability [58] showed that smaller diameter tubes were oxidized more rapidly than larger diameter tubes. The finding is consistent with our data for a series of (n,n) and (m,0) SWCNTs highlighting quantitatively decreasing the tube chemical susceptibility at increasing tube diameters.

3. Frequently intrinsic defects on SWCNTs are supplemented by oxidative damage to the nanotubes framework by strong acids which leave holes functionalized with oxygenated groups such as carboxylic acid, ketone, alcohol, and ester group [59]. This and other studies (see reviews [6, 9, 60]) point to high chemical reactivity of the defects, that is in good agreement with our computed data.

4. Attributing endcaps of SWCNTs to the chemically active regions is widely accepted [4, 61]. However, as shown by calculations, this region is not always the most active since its activity can be overcome by a particular composition of the open ends.

5. There was a skepticism concerning the sidewall activity, which was considered much smaller with respect to that of fullerenes [4-15]. However producing small diameter SWCNTs has removed the doubts opening the way to large scale hydrogenation [62], fluorination [63], amination [64], and a great number of other addition reactions [10, 65] involving SWCNTs sidewalls.

Therefore, available chemical data well correlate with predictions to be made on the basis of the computations performed. Particularly, the available data on experimental study of hydrogenated SWCNTs [62] should be noted. As shown, a significant perturbation of SWCNTs structure under hydrogenation occurs. The finding seems to be understood on the basis of our data concerning the tube with both H-terminated open ends (see Fig. 7). The computations clearly highlight that attaching hydrogen atoms to the most active places of the tube causes a significant disturbance of its $N_{DA}$ map and, consequently, the tube structure which is spread over large region. Obviously, the more atoms are attached, the more drastic changes occur that is observed experimentally.

## 4.3. ELECTRONIC CHARACTERISTICS

The studied SWCNT fragments provides a large set of data summarized for (4,4) fragments in Table 2. The analysis of the main electronic characteristics presented in the table allows for making the following conclusions.

1. According to the fragments energy, endcap/H-terminated-open-end and H-terminated-open-end/H-terminated-open-end tubes are the most energetically stable.

2. Removing eight hydrogen terminators from endcap/H-terminated-open-end tubes causes energetic lost of 465.7 and 486.3 kcal/mol in the case of shorter and longer tube, respectively. Those figures correspond to -58.2 and -60.8 kcal/mol (or -2.5 and -2.6 eV) which accompany the formation of a single C-H bond in these cases. The data are in good consistence with both experimental [62] and calculated data [14, 15] estimating the energy of the formation of one C-H bond of ~2.5eV.

3. Substitution of eight terminators by the second cap requires energy of 264.8 kcal/mol.

4. Emptying one open end of the H-terminated-open-end/H-terminated-open-end fragment costs of 418.2 kcal/mol ( -52.3 kcal/mol per one C-H bond) while additional removing eight hydrogen terminators from the other end adds 447.3 kcal/mol (-55.9 kcal/mol) to the energetic lost. As seen, the obtained data for endcap/open-end tube and open-end/open-end tube well coincide.

5. All endcap/H-terminated-open-end tubes are polarized with large dipole moment of ~10 D while endcap/empty-open-end and endcap/endcap tubes have very low dipole moment if nothing. Practically, this circumstance is very important when applying SWCNTs either to improve characteristics of nonlinear optical devices based on liquid crystalline media [65, 66] or to design new hybrid materials based on SWCNTs and electron donor-acceptor nanocomposites [10].

6. Among two-open-end tubes, the H-terminated-open-end/empty-open-end tube is the most polarized while H-terminated-open-end/H-terminated-open-end and empty-open-end/empty-open-end tubes have small dipole moments.

7. Important to note a high level of donor-acceptor characteristics of the tubes. All tubes are characterized by not too high ionization potentials and high electron affinity. The characteristic value $I_D$-$\varepsilon_A$ is in interval of 7.4-5.9 eV that meets the requirements of the formation of tightly coupled adducts involving two or more SWCNTs leading to their donor-acceptor stimulated adhesion similarly to the dimerization of fullerene $C_{60}$ [67]. At the same time this explains a high acceptor ability of SWCNTs in numerous donor-acceptor nanocomposites [10].

Once related to SWCNTs of the (4,4) family, the obtained regularities can be nevertheless spread over other (n,n) and (m,0) tubes due to close similarity in their behavior exhibited in the current study. Taking together, the obtained results clearly show a high efficiency of UBS HF for detailed quantitative characterization of SWCNTs chemical reactivity.

**ACKNOWLEDGMENTS**. The work was supported by the Russian Foundation for Basic Research (grant 08-02-01096).

**Table 1**. Atomic chemical susceptibility of H-terminated nanographenes

| Nanographenes ($n_a, n_z$)[1] | $N_{DA}$ | | |
|---|---|---|---|
| | Armchair edge | Central part | Zigzag edge |
| (15, 12) | 0.28-0.14 | 0.25-0.06 | 0.52-0.28 |
| (15, 12)[2] | 1.18-0.75 | 0.25-0.08 | 1.56-0.93 |
| (7, 7) | 0.27-0.18 | 0.24-0.12 | 0.41-0.28 |
| (5, 6) | 0.27-0.16 | 0.23-0.08 | 0.51-0.21 |

[1] Following [32], $n_a$ and $n_z$ match the numbers of benzenoid units on the armchair and zigzag ends of the sheets, respectively.
[2] After removing hydrogen terminators

**Table 2**. Comparison of bond dissociation energy (BDE, in eV) of zigzag edge-X bonds with experimental BDE of $C_2H_5$-X. Zero-point-vibration energy corrections are not included

| Radical X | H | OH | $CH_3$ | F | Cl | Br | I |
|---|---|---|---|---|---|---|---|
| BDE (edge X) UBS DFT[1] | 2.86 | 2.76 | 2.22 | 3.71 | 2.18 | 1.65 | 1.18 |
| BDE (edge X)[2] UBS HF | 4.27 | 3.86 | 3.42 | 4.50 | 2.25 | - | - |
| BDE ($C_2H_5$-X)[3] | 4.358 | 4.055 | 3.838 | 4.904 | 3.651 | 3.036 | 2.420 |

[1] Data from Ref.[41]
[2] The current study calculations were performed for NGR (**5, 6**).
[3] Experimental values from Ref. [42].

**Table 3.** NGrs electronic characteristics in *kcal/mol*

| Nanographenes[1] | The number of "magnetic"(odd) electrons | $E_{S=0}^{UBSHF}$ | $J$ | $E_{S=0}^{PS}$ | Singlet-triplet gap[2] |
|---|---|---|---|---|---|
| (15, 12) | 400 | 1426.14 | -0.42 | 1342.14 | 0.84 |
| (7, 7) | 120 | 508.69 | -1.35 | 427.69 | 2.70 |
| (5, 6) | 78 | 341.01 | -2.01 | 262.72 | 4.02 |

[1]Nomenclature of nanographenes is given in Footnote 2 to Table 1.
[2]For pure-spin states the singlet-triplet gap $E_{S=1}^{PS} - E_{S=0}^{PS} = -2J$ [16].

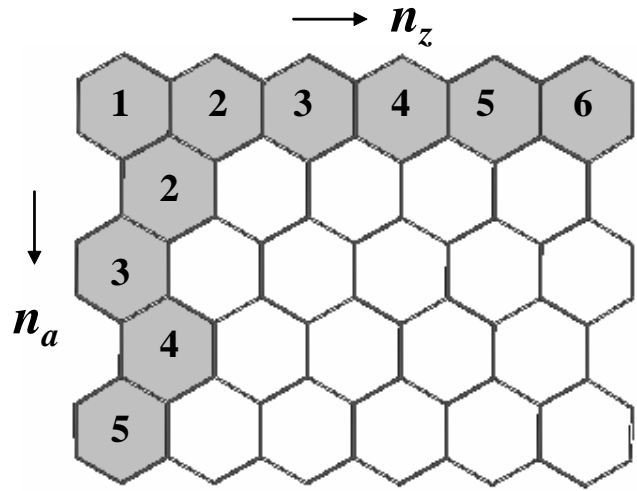

**Fig. 1.** Notification of a rectangular NGR ($n_a$. $n_z$) with armchair ($n_a$) and zigzag ($n_z$) edges [32]

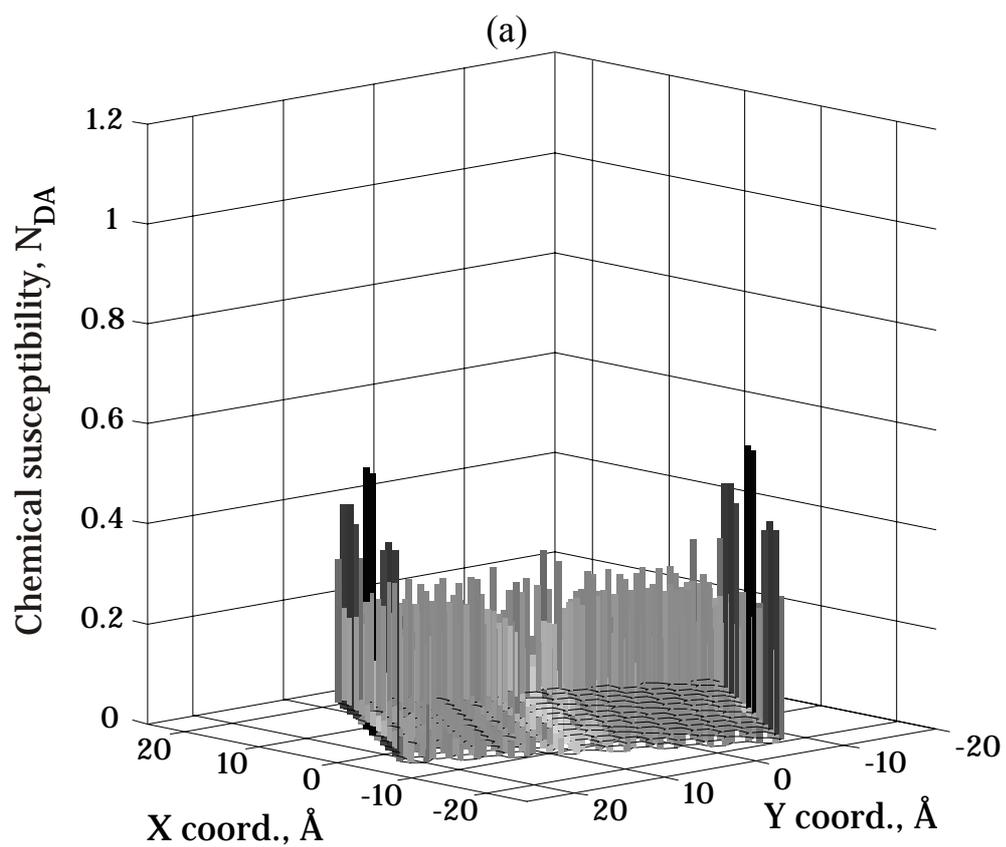

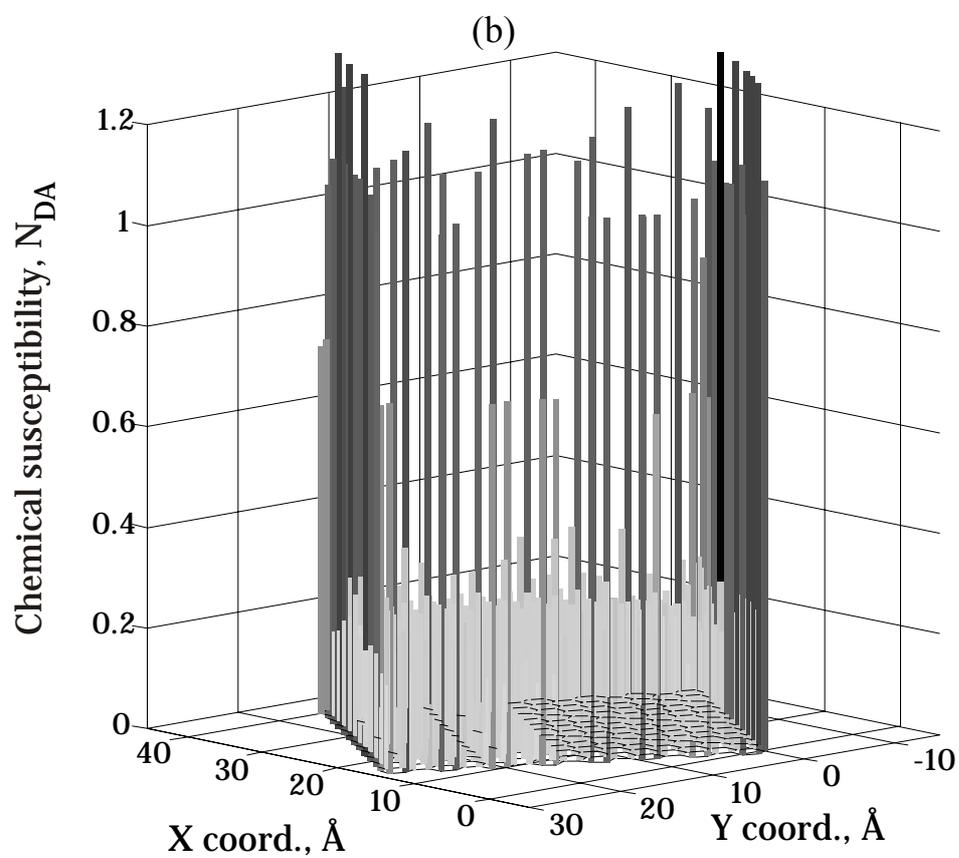

**Fig.2.** Distribution of atomic chemical susceptibility over atom of rectangular NGR (**15,12**) with hydrogen terminated (a) and empty (b) edges. UBS HF solution. Singlet state.

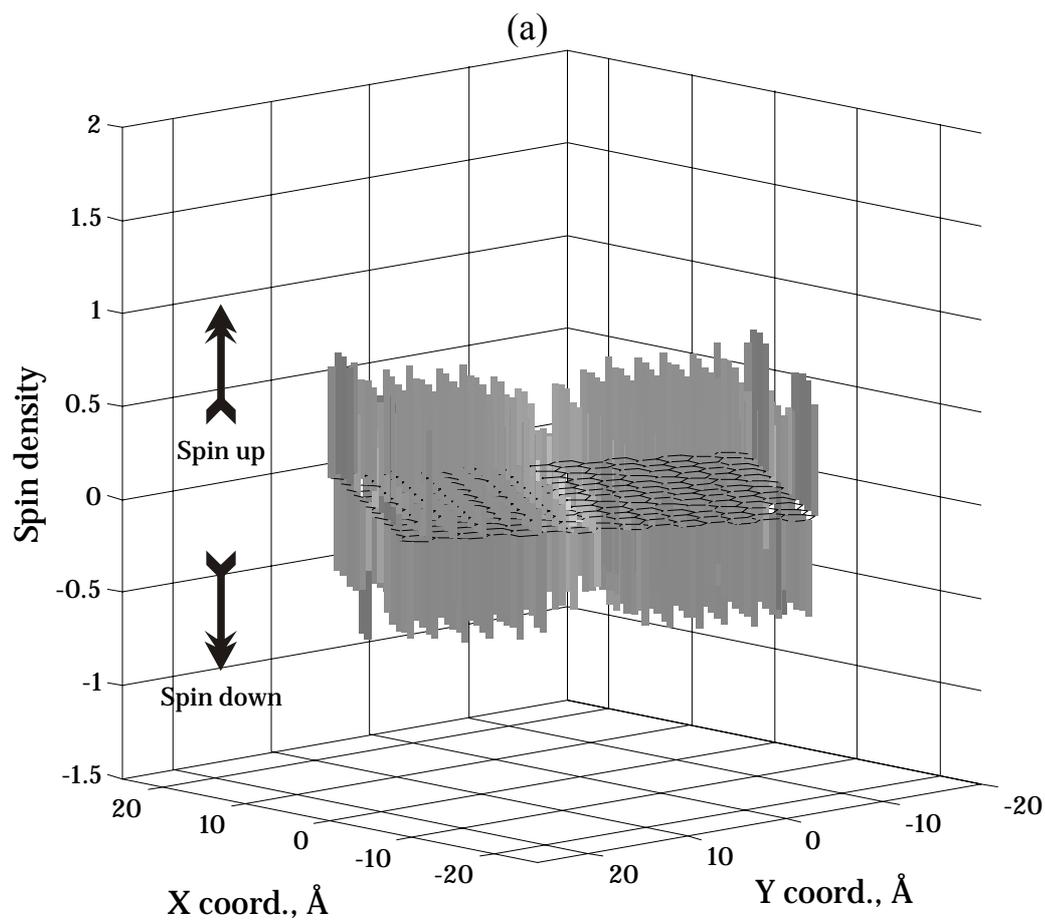

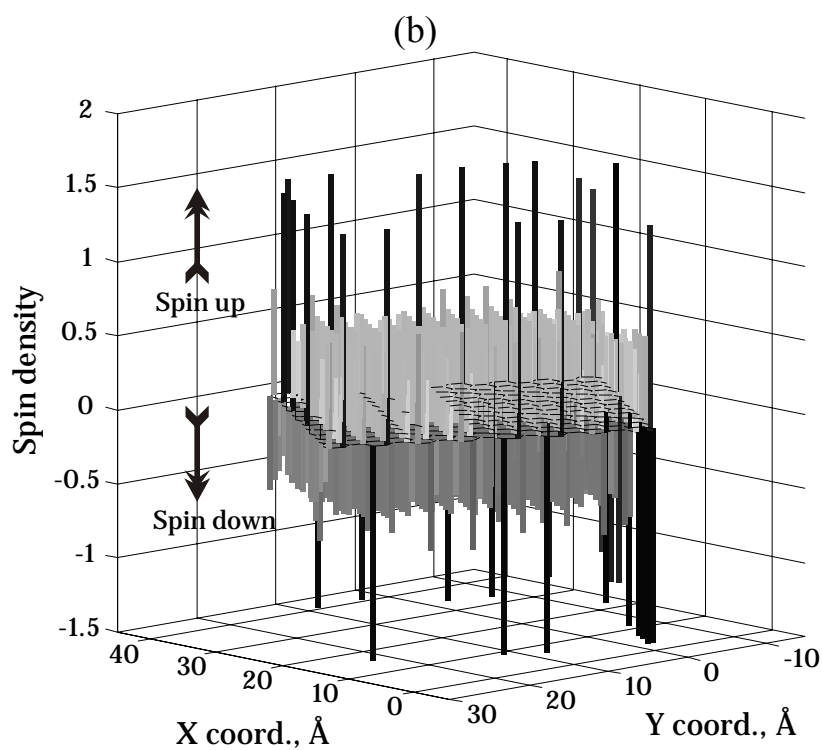

**Fig.3.** Distribution of spin density over atom of rectangular NGR (**15,12**) with hydrogen terminated (a) and empty (b) edges. UBS HF solution. Singlet state.